\documentstyle[epsf,12pt]{article}
\newcommand{\al}{\alpha'}
\newcommand{\de}{\partial}
\newcommand{\be}{\begin{equation}}
\newcommand{\ba}{\begin{eqnarray}}
\newcommand{\ea}{\end{eqnarray}}
\newcommand{\ee}{\end{equation}}

\newcommand{\f}{\frac}
\newcommand{\s}{\sqrt}
\newcommand{\vp}{\varphi}

\newcommand{\ap}{\alpha}

\newcommand{\ddd}{\cdot\cdot\cdot}

\newcommand{\la}{\langle}
\newcommand{\lb}{\rangle}
\newcommand{\ep}{\epsilon}

\newcommand{\dz}{\mbox{D}0-\overline{\mbox{D}0}}

\begin{document}
\begin{titlepage}
\thispagestyle{empty}
\begin{flushright}
HUTP-05/A0013 \\
hep-th/0503184 \\
March, 2005
\end{flushright}

\bigskip

\begin{center}
\noindent{\Large \textbf{
$c=1$ Matrix Model from String Field Theory
}}\\
\vspace{2cm} \noindent{Tadashi
Takayanagi\footnote{takayana@bose.harvard.edu} and Seiji
Terashima\footnote{seijit@physics.rutgers.edu}}\\
\vspace{1cm}

${}^1$
{\it Jefferson Physical Laboratory,  Harvard University \\
Cambridge, MA 02138, USA}

\vspace{3mm}
 ${}^2$
 {\it  New High Energy Theory Center, Rutgers University\\
126 Frelinghuysen Road, Piscataway, NJ 08854-8019, USA}

\end{center}
\begin{abstract}

We show that
the boundary string field theory (BSFT)
on unstable D$0$-branes in 2d string
theory is equivalent to the double scaled $c=1$ matrix
model (i.e. quadratic action),
even though we naively expect many interaction terms in BSFT.
It is checked that S-matrices are trivial in the
open string theory on the D$0$-branes.
We discuss how to take the closed string decoupling limit in our
model in order to make its holographic interpretation clear.
We also find some useful lessons from our
results in 2d string toward the further understanding of the
higher dimensional BSFT action.

\end{abstract}
\end{titlepage}

\newpage

\tableofcontents

\section{Introduction}

It is important to obtain
a non-perturbative and background independent
formulation of the string theory
in general backgrounds in order to
 understand some deep problems, for example,
the resolution of the cosmological singularities in string theory.
A significant progress in this direction is the BFSS
matrix model proposal \cite{BFSS}.
The action of the BFSS matrix model is the
same as a low energy effective action on multiple D$0$-branes in
type IIA superstring. However, this model is supposed to describe
M-theory in an infinitely boosted flame. Another related problem is
that a D$0$-brane in type IIA superstring has a conserved charge,
i.e. D$0$-charge, so we can not construct D-branes without the
D$0$-charge. This means that the
D-branes always need to have nonzero field
strengths and thus non-commutative world-volumes.

In order to overcome this difficulty,
one might consider the matrix model based on
non-BPS D$0$-branes or ${\mbox{D}0-\overline{\mbox{D}0}}$ pairs
\cite{K-matrix}
since these branes have no conserved charges
and it was shown that
we can construct any D-branes from them \cite{Te, K-matrix}.
However, because of the tachyons in these unstable D$0$-branes
we should employ a string field theory to analyze
the dynamics of the unstable D$0$-branes. Since it includes infinitely
many massive modes and derivatives,
such a matrix model
does not look tractable as it is.

Recently, it was proposed \cite{MV} that the $c=1$ matrix model,
which had been known to describe the two-dimensional (2d) string
theory \cite{twodmat}, can be considered as a tachyon action on
multiple D$0$-branes. This matrix model description of 2d closed
string theory is argued to come from a open-closed duality. Soon
later, these D$0$-branes are identified with the so called
ZZ-branes \cite{ZZ} in the boundary Liouville theory
\cite{KMS,MVT,Senreview}. This idea was also successfully applied
to the construction of a matrix model for the two dimensional type
0 string \cite{TT,six}.

Thus if this proposal is completely correct, we can seriously
regard the $c=1$ matrix model as a concrete realization of the
idea of the matrix model based on unstable D$0$-branes. Even
though it has been established that a fermion in the double scaled
$c=1$ matrix model is identified with a decaying unstable
D$0$-brane, we only have a few evidences for the identification of the
$c=1$ matrix model with a tachyon action \cite{SenOC} \cite{KMS}.
Though we know that the
perturbative tachyon mass agrees with that of the $c=1$ matrix
model \cite{KMS}, we are lacking precise
understanding of the tachyon action
of ZZ-branes including interactions.

Naively, it is naturally expected that there are non-trivial
interaction terms as is true for string field theory actions in
higher dimension. Then we immediately encounter a sharp paradox
why the double scaled $c=1$ matrix model is so simple that it does
not include any interaction terms. One may think that taking the
double scaling limit may save the situation since it might reduce
the complicated tachyon action to the simplified quadratic action
as suggested in \cite{MV}. However, this is problematic as long as
we consider putting D$0$-branes in the conventional 2d string theory
defined by the Liouville theory. This is because when we put an
extra D$0$-brane as a probe, its action from the matrix model
disagrees with the one from open string field theory.

We resolve this problem by considering the boundary string field
theory (BSFT) action \cite{BSFT,GeSh,KMM,KMM2,KL,TTU,ZS} on the
ZZ-branes. The BSFT action for unstable D$0$-branes is originally a
complicated one including infinitely many higher derivative terms.
However, it can be greatly simplified via a field redefinition.
Indeed, in the present paper we will argue that it is
(classically) equivalent to the quadratic free action, which is
exactly the same as the double scaled $c=1$ matrix model.

We will give several independent evidences for this claim. First,
we directly check that the on-shell four point interaction is
vanishing in the open string theory on a ZZ-brane. Next, we extend this
result to arbitrary interactions at any order of the perturbation
theory by applying a slightly indirect approach of taking $c=1$
limit of the non-minimal $c<1$ string \cite{TLD}. Finally, we move
on to a direct examination of the field redefinition which
relates the BSFT action and the simple quadratic action. We
explicitly construct the field redefinition up to the cubic order
in the BSFT action and check that it is indeed non-singular.

The organization of this paper is as follows. We first state our
main claim that the BSFT action on unstable D$0$-branes in 2d string
theory is classically equivalent to the double scaled $c=1$ matrix
model (i.e. quadratic action) in section two. Then we give
evidences for this claim in section three by analyzing the
on-shell amplitudes and the BSFT action. In section four we
discuss how to take the closed string decoupling limit in our
model in order to make its holographic interpretation clear. In
the section five we try to find some useful lessons from our
results in 2d string toward the further understanding of the
higher dimensional BSFT action. In section six we draw conclusions
and discuss future problems.


\section{Main Claim: $c=1$ Matrix Model from BSFT action}

The main claim we would like to argue in this paper is that {\it
at the the classical level (or tree level), the BSFT action for
non-BPS $N$ D$0$-branes (ZZ-branes in the Liouville theory) in two
dimensional bosonic or type0 string
 is equivalent to the quadratic action} \be
S=\frac{1}{g_s} \int dt {\rm Tr} \left[\f{1}{2}(D_{t}\Phi)^2+\f{1}{2}\Phi^2-1
\right],\label{cmat} \ee
 where $\Phi$ is a $N$ by $N$ hermitian matrix
and $D_{t}=\de_{t}-i[A_0,]$
is the covariant derivative of the $U(N)$ gauge symmetry.\footnote{
We have set $\alpha'=1$ for bosonic string and 
$\alpha'=\f{1}{2}$ for type 0 string.}
Of course, if we take a limit $N \rightarrow \infty$, this action is
same as the double scaled $c=1$ matrix model action except the
constant energy shift. We can also obtain the BSFT action for $N$
$\dz$ branes just by replacing the Hermitian matrix $\Phi$ and the
gauge group $U(N)$ with a complex matrix and $U(N)\times U(N)$,
respectively. A closely related proposal for general string field
theories in 2d string was also proposed in \cite{SenOC}.

In other words, we can relate both theories via a non-singular
field redefinition. At a higher loop level, we need to take the
measure of path-integral into account and this issue is beyond the
scope of the present paper. However, as we will discuss later in
the next section, we have a strong evidence which supports the
action (\ref{cmat}) is correct even if we include all perturbative
loop corrections\footnote{Clearly this agrees with what we expect
from the $c=1$ matrix model.}.

It is recently proposed that the $c=1$ matrix model can be
identified with the theory of unstable D$0$-branes and it describes
the closed string theory via a kind of open-closed duality or
holography \cite{MV}. In 2d string theory, an unstable D$0$-brane
can be described by a ZZ-brane \cite{ZZ} in the Liouville CFT with
the Neumann boundary condition in the time-direction \cite{KMS}.
The physical field on it is only the tachyon field $T$ and the
gauge field $A_{0}$. There is no transverse scalar field because
the linear dilaton prevents any movements of the D-brane in the
Liouville direction. Thus, the open string theory on the unstable
D$0$-branes agrees with the $c=1$ matrix model {\it when we neglect
interactions}. To see if this idea does really make sense, we need
to understand the reason why there are no interaction terms like
$\Phi^3,\Phi^4$ etc. in the (double scaled) $c=1$ matrix model
(\ref{cmat}).

Our claim, presented just before, not only supports this
conjecture, but also shows that the interaction terms in BSFT can
be set to zero via the field redefinition. Consider $M$ D$0$-branes
in 2d string theory. In the $c=1$ matrix model description, we can
regard them as $M$ extra eigenvalues $x_i\ \ (i=1,2,\ddd,M)$
sitting above the Fermi level\footnote{In spite the energy shift
in (\ref{cmat}), below we define the energy for each eigen-value
by the standard formula $H=\f{1}{2}(p^2-x^2)$. Here we have used a
rescaled coordinate $x = \Phi \sqrt{g_s}$ and its conjugate
momentum $p$.} $E=-\mu$. It is well-known that all eigenvalues in
the matrix model behave as free fermions. When there is no open
string tachyon condensation on them $T=0$, they are sitting at the
top of the potential $x_i=0$. Then the $M$ D$0$-branes can be
described by the matrix model (\ref{cmat}) with the $M\times M$
matrix $\Phi$. Since we cannot put a fermion below the Fermi
level, we must require the important constraint \be
p^2-x^2\geq-2\mu. \label{cons} \ee On the other hand, when we
consider the BSFT action for the $M$ unstable D$0$-branes, our claim
shows the classical equivalence between the $c=1$ matrix model
action and the BSFT action via a field redefinition.

The constraint (\ref{cons}) intuitively corresponds to the fact
that the tachyon potential\footnote{Interestingly, the exponential
potential in BSFT is directly related to the loop operator which
creates another kind of D-brane (so called FZZT-brane) as noticed
in \cite{TDB}. The relation between this observation and our
arguments in this paper is not obvious at present.}
 in BSFT is bounded from
below. In bosonic string, the potential is given by
$V(T)=(T+1)e^{-T}$ \cite{KMM}. In this case this potential is
bounded for the positive $T$, while unbounded for the negative
$T$. This nicely agrees with the $c=1$ matrix model description of
2d bosonic string \cite{twodmat}, where only one side is filled
with the fermions. On the other hand, in the superstring case, the
potential of non-BPS D$0$-branes and $\dz$ branes are given by the
Gaussian profile $V(T)=e^{-T^2}$ \cite{KMM2} and
$V(T,\bar{T})=e^{-T\bar{T}}$ \cite{KL,TTU}, respectively. Since
they are completely bounded from below, it again matches with the
fermi surface structures in the
type 0B and 0A matrix model \cite{TT,six}.

When the tachyon $T$ is not a constant, its profile with
$E(T,\dot{T},\cdots)<-\mu$ at an instant should be excluded. If we
compare the classical solutions in both sides, the constraint
(\ref{cons}) in the $c=1$ matrix model matches with that of the
$\lambda$ parameter
$0\leq \lambda\leq 1/2$
\cite{originalroll}.
Note that the classical solutions of BSFT should be the same
as those of BCFT or the boundary state
because they all use the same boundary perturbation
and are essentially the same for the on-shell structure.
In this way, the claim is consistent with the
open-closed duality interpretation of $c=1$ matrix model in a
highly nontrivial way. We will see more direct evidences in the
next section.

\section{Evidences for the Equivalence}

In this section we will give important evidences for the claim
stated in the previous section.

First we study on-shell amplitudes of open string tachyons
in 2d string theory.
It is very important to notice that the momenta of the
on-shell tachyons are severely restricted from the energy-momentum
conservation in one dimensional D-brane world volume. Actually,
this requires that the in-state
and out-state should be the same because there are
only two on-shell states ($e^{X^0/\s{\al}}$ and
$e^{-X^0/\s{\al}}$).

 However, we still have the possibility that
some nontrivial phase for the S-matrix elements exists like the
scalar $\phi^4$ theory. Below, we will argue that in the ZZ-brane
case the S-matrix is indeed trivial, i.e. $S=1$. We show this by
an explicit computation of the four point interaction at tree
level in section 3.1, and also by a slightly indirect approach at
any order of the perturbation in section 3.2. Since any smooth
field redefinition does not change S-matrices, we expect they are
trivial on the D$0$-branes in 2d string theory if their action is
indeed equivalent to the quadratic one of the $c=1$ matrix
model.\footnote{Strictly speaking, this S-matrix we consider is
formal because we can not define the asymptotic field in
one-dimensional "field theory". However, this formality of the
discussion is sufficient for showing the equivalence because we do
not want to consider the physical scattering amplitude.} It is
noted here that that the BSFT is supposed to reproduce the
perturbative S-matrix. This is because the BSFT action is reduced
to the partition function when all fields are on-shell and it can
be regarded as S-matrix generating effective action
\cite{AndreevTseytlin,Tseysigma,HaTe}.

Next we study the BSFT action itself order by order with respect
to the power of the tachyon field. We explicitly compute the field
redefinition which relates the BSFT action to the $c=1$ matrix
model up to the cubic order
and show that it is indeed non-singular in section 3.3.

\subsection{Four Tachyon on-shell Scattering Amplitude}
\setcounter{equation}{0}

 Since the scattering amplitude
 for odd number tachyons are all zero
due to the momentum conservation, the lowest non-trivial one is
the four tachyon amplitude. Thus let us compute this and check
that it is vanishing. The Liouville sector has no effect in this
computation because only the time coordinate $X^0$ obeys the
Neumann boundary condition on the D$0$-brane. Therefore we can just
reduce the standard formula of S-matrices in $D=26$ (or $D=10$)
dimensional (super)string to those in $D=1$ dimension.
 For general dimension $D$, the well-known
result for bosonic string (i.e. Veneziano amplitude, see e.g.
\cite{PO} ) is given by setting $\al=1$ \ba
&&S(k_1,k_2,k_3,k_4)=2i g_0^2 (2\pi)^D \delta^D(\sum_i k_i) \\
&&\ \ \times \left[\f{\Gamma(-s-1)\Gamma(-t-1)}{\Gamma(-s-t-2)}+
\f{\Gamma(-t-1)\Gamma(-u-1)}{\Gamma(-t-u-2)}+
\f{\Gamma(-u-1)\Gamma(-s-1)}{\Gamma(-u-s-2)}\right],
\label{amppf} \ea
where we have defined the Mandelstam variables $s=-(k_1+k_2)^2,\
t=-(k_1+k_3)^2$ and $u=-(k_1+k_4)^2$. The on-shell condition for a
tachyon vertex operator $e^{ikX}$ is $k^2=1$. The variables $s,t$
and $u$ are not independent and obey the constraint $s+t+u=-4$ in
any dimension $D$. In the one dimensional case, which we are
interested in, we can choose the values
$k_1=k_2=1$ (incoming) and $k_3=k_4=-1$
(outgoing), or equivalently $s=-4$ and $t=u=0$. At this point,
however, the amplitude (\ref{amppf}) becomes singular. To
regularize it, we first assume $D>1$ and finally take the limit
$D=1$. We define the infinitesimal variable $\ep$ by $s=-4+\ep$
and also assume $t$ and $u$ are infinitesimal of the same order.
Then the sum of the $\Gamma$ function ratios in (\ref{amppf}) can
be written as follows up to $O(\ep)$ terms\footnote{To show this
we used the formula $\Gamma(-n+\ep)=\f{(-1)^n}{n!}(\epsilon^{-1}+
(\sum_{k=1}^n k^{-1})-\gamma)+O(\ep)$ for a zero or positive
integer $n$.}
 \be
\f{\Gamma(-s-1)\Gamma(-t-1)}{\Gamma(-s-t-2)}+\ddd
=-\f{(t+u)(\ep+t+u)}{tu}=0. \ee In the final equality we have used
the constraint $s+t+u=-4$.

We can also do the similar computations\footnote{ In this case the
amplitude is proportional to
$\f{\Gamma(-s)\Gamma(-t)}{\Gamma(-s-t-1)}+(2$ permutations), when
we set $\al=1$. The kinematical constraint is $s=-2,t=u=0$. We can
check this is vanishing in the almost same way as in the bosonic
string.
} in the
superstring (i.e. type 0 string) case and find the same
conclusion.

This absence of the four tachyon interaction may be understood
from the fact that the Veneziano amplitude is
a sum of the contributions from
the propagators of open string modes.
Because there are no massive or massless modes
on the ZZ-brane,
it is natural that the four point interaction vanishes
for the ZZ-brane.

Thus we have found that the four particle scattering amplitude in
the open string theory is vanishing. Then it is natural to guess
that all non-trivial on-shell scattering amplitudes are zero at
tree level.

\subsection{General On-shell Amplitudes via $c<1$ Deformation}

To avoid the singular behavior in the $c=1$ string (=2d string),
it is also helpful to assume an infinitesimal background linear
dilaton in the time-direction in addition to the standard
space-like linear dilaton. Actually this was the way how the
closed string S-matrices were computed in 2d string theory
\cite{DK}. Then the coupling constant behaves like
$g_s=e^{qX^0+Q\phi}$, where we defined $q=1/b-b$ and $Q=1/b+b$.
This theory describes a non-critical string with a matter central
charge $c=1-6q^2<1$ (see \cite{TLD}). We also need to add the
Liouville term $\mu \int dz^2 e^{2b\phi}$ to regulate the strongly
coupled region as usual. The matrix model dual of this $c<1$
string was constructed in \cite{TLD} and it is given by the action
(\ref{cmat}) with the constant string coupling $g_s$ replaced with
the time-dependent one $g_s\cdot e^{qt}$. In the final stage, we
will take the limit $b\to 1$, which is equivalent to the $c=1$
string. The analogous discussion in type 0 string can be done in
the same way and we will omit it.

Since only the time coordinate $X^0$ obeys the Neumann boundary
condition, there are only two possible choices (in-coming and
out-going states) for the physical vertex operator. They are given
by  \be V_{-}=e^{-bX^0},\ \  \mbox{and }\ \ V_{+}=e^{X^0/b}. \ee

Now let us consider the on-shell scattering amplitudes by using
these vertex operators. A crucial condition is the energy
conservation and we have to check this first. Notice that there is
no Liouville term in the time direction. Thus we get the
constraint from the energy conservation (this is more obvious when
we assume the Euclidean time) \be
-N_-b+\f{N_+}{b}=\left(\f{1}{b}-b\right)\chi \ \ , \label{cons2}
\ee where $N_{\pm}(\geq 0)$ denotes the number of the insertions of
$V_{\pm}$, and $\chi$ is the Euler number.

Let us assume that $b^2$ is an irrational number. We can rewrite
(\ref{cons2}) as \be N_+-N_-b^2=\chi-\chi b^2. \ee This
immediately requires \be N_{+}=N_-=\chi. \ee When we are
interested in the disk amplitudes ($\chi=1$), only two point
function $N_+=N_-=1$ satisfies this condition. Also for the
cylinder amplitudes ($\chi=0$) only zero point function is
non-trivial. All higher open string amplitudes should be zero
since they do not satisfy the energy conservation. Thus $N\geq 3$
particle on-shell scatterings are all zero at any order of
perturbation theory.

Next we require that the amplitudes should be continuous with
respect to the parameter $b$. This is also expected from its
matrix model dual \cite{TLD}. Then we can obtain the same
conclusion for any values of $b$. Finally, taking the limit $b=1$,
we have the desired conclusion that
 the $N\geq 3$ particle on-shell scatterings are
all zero at any order of perturbation theory in the 2d string
theory.

If we consider a theory where we cannot change the value of $b$
continuously, it is possible that some of non-trivial amplitudes
are non-zero only when $b^2$ is a rational number (see
(\ref{cons2})). This seems to correspond to the minimal string
case \cite{minimal} (see also e.g. \cite{SS} for modern
viewpoints). Indeed, in minimal models it is impossible to change
the values of $b$ continuously as the $(p,q)$ minimal model
corresponds to the value $b=\s{p/q}$.

In this way, we have seen that open string tachyons for unstable
D-branes have no interaction with themselves in on-shell
amplitudes. This conclusion strongly supports our claim in section
2 that the open string field theory is equivalent to the free
quadratic action via a smooth field redefinition. This is because
any smooth field redefinition does not change S-matrices.

\subsection{Field Redefinition between BSFT and Matrix Model}

Here we would like to find a field redefinition which relates the
BSFT action to the $c=1$ matrix model one up to the cubic order.
We will show that it is indeed smooth one\footnote{
Here smooth means smooth near the on-shell momentum.}.
We neglect open string
modes other than the open string tachyon into the BSFT action of
the ZZ-brane since they are expected to play no important role in
BSFT at the classical level.

We consider the open string theory on an unstable D$0$-brane in 2d
string theory. The important point is that this theory is again
equivalent to a open string theory in one dimension because open
string modes on a ZZ-brane do not excite the Liouville field
$\phi$. Thus the BSFT action for a D$0$-brane in 2d string is the
same as the zero dimensional truncation of the BSFT action in the
$26$ dimensional critical string or,
if we consider type 0 string theory, the one
in the $10$ dimensional critical string theory.
Then we can also generalize the
same result to the case of multiple D$0$-branes because we can
always diagonalize the matrix $T$ by the $U(M)$ gauge fixing,
leading to essentially the Abelian action.

The BSFT action in bosonic string up to the cubic order was
computed in \cite{Coletti, Frolov}
as
follows after the Fourier transformation

\ba S_{BSFT}&=&\frac{1}{g_s}
\Bigl[2 \pi \delta(0)-\f{1}{2}\int dk(2\pi)A(k)T(k)T(-k)
\\ &+&\f{1}{6} \int
dk_1dk_2(2\pi)B(k_1,k_2,-k_1-k_2)T(k_1)T(k_2)T(-k_1-k_2)+\ddd
\Bigr], \label{bsftac} \ea
where we introduced two functions $A(k)$ and
$B(k_1,k_2,k_3)$ defined by
\ba && A(k)=\f{\Gamma(2-2k^2)}{\Gamma(1-k^2)^2}
=2 \f{4^{-k^2}}{\sqrt{\pi}} \f{\Gamma(\f{3}{2}-k^2)}{\Gamma(1-k^2)},
\nonumber \\
&& B(k_1,k_2,k_3) =2(1+\sum_{i<j}k_ik_j)I(k_1,k_2,k_3)
-(J(k_1,k_2,k_3) +\mbox{cyclic permutations}).
\nonumber
\ea 
Here $2 \pi \delta(0)=\int dt$ and we defined
\ba
2 (1+\sum_{i<j}k_ik_j) I(k_1,k_2,k_3) &=&
\f{4^{\sum \alpha_i}}{4 \pi}
 \f{ \Gamma(\sum \alpha_i+1/2) }{\sqrt{\pi}}
\f{ \Pi_{i=1}^3  \Gamma(\alpha_i) }
{ \Pi_{j<k}^3 \Gamma(\alpha_j+\alpha_k) }, \nonumber \\
J(k_1,k_2,k_3) &=&
\f{4^{\sum \alpha_i}}{4 \pi}
\f{\Gamma(\alpha_2+\alpha_3+1/2)}
{\Gamma(\alpha_1+1/2)}
\f{\Gamma(\alpha_1) }{\Gamma(\alpha_2+\alpha_3)},
\label{IJ}
\ea
where $\alpha_1=k_2k_3+1/2, \alpha_2=k_1k_3+1/2,
\alpha_3=k_1k_2+1/2$, which becomes $\alpha_1=\alpha_2=\alpha_3=0$
when all three momenta are on-shell.


Then we can find a field redefinition into the $c=1$ matrix model
\be T(k)=\alpha(k)\Phi(k)+\int
dk'\beta(-k,k')\Phi(k-k')\Phi(k')+\ddd, \ee where we defined \ba
&& \alpha(k)=\s{\f{1-k^2}{A(k)}},\\ && \beta(k_1,k_2)= \Biggr[
\f{\alpha(k_2)\alpha(k_3)}{2A(k_1)} \left( \f{2}{3}(1+
\sum_{i<j}k_ik_j         )I(k_1,k_2,k_3)
-J(k_1,k_2,k_3)\right) \Biggl]_{k_3=-k_1-k_2}. \label{fieldredef}
\ea By this field redefinition, the BSFT action (\ref{bsftac}) is
actually mapped to the matrix model action (\ref{cmat}). Here we
have used the fact that the cyclic permutations of $J$ give the
same contribution in (\ref{bsftac}). It is clear that the factor
$\ap(k)$ is non-singular since $A(k)$ has a zero at the on-shell
point $k^2=1$. Next, another one $\beta(k_1,k_2)$ looks divergent
at $k_1^2=1$ since it includes the factor $A(k_1)^{-1}$. However,
we can see that it is successfully canceled with the zero of the
other factor expressed by the $1/\Gamma(\alpha_2+\alpha_3) \sim
(k_1^2-1)$ in (\ref{IJ}). \footnote{ In other words, the cubic
term is proportional to the on-shell factor $k^2-1$. This property
is very special for the BSFT action and also important for a
consistency check of the BSFT as a string field theory (the
vanishing of the 1-particle reducible diagram) \cite{Frolov}
\cite{HaTe}.} Thus we can conclude this field redefinition is
non-singular.

One might think this can also be applies to higher dimensional
D-branes. However, this is not true. As pointed out in
\cite{HaTe}, near the on-shell point $|\alpha_i| <<1$, we find
\begin{eqnarray}
B(k_1,k_2,k_3)=
\f{1}{4\pi} \left(
\f{ \Pi_{j<k}^3 \alpha_j+\alpha_k }{ \Pi_{i=1}^3  \alpha_i }
-\f{\alpha_1+\alpha_2}{\alpha_3}-\f{\alpha_2+\alpha_3}{\alpha_1}
-\f{\alpha_3+\alpha_1}{\alpha_2} \right) +{\cal O}(\alpha_i)
=\f{1}{2\pi} +{\cal O}(\alpha_i), \nonumber
\end{eqnarray}
where we assumed all $\alpha_i$ are of the same order. This
$\frac{1}{2\pi}$ is the correct S-matrix for three open string
tachyons. Thus the BSFT reproduces the on-shell interaction term
for three open string tachyons in a highly nontrivial way.
Therefore, the field redefinition (\ref{fieldredef}) becomes
singular when $k$ is on-shell because the momenta $k_1,k_2$ and
$-k_1-k_2$ can be on-shell at the same time. In one-dimension, it is
impossible that all three momenta are on-shell due to the momentum
conservation. This is crucial for the smoothness of the field
redefinition.

We note that this field redefinition is also consistent with the
rolling tachyon solution \cite{originalroll}. The BCFT solution
$T=\widetilde{\lambda} \cosh (t) $ should correspond to the
solution $\Phi=\sin (\pi \widetilde{\lambda}) \cosh(t)$ in the
matrix model by matching the energy \cite{KMS,Senreview}. For the
profile $\Phi=\sin (\pi \widetilde{\lambda}) \cosh(t)$, we obtain
$\alpha(k)=1$, and the quadratic term in (\ref{fieldredef})
vanishes since $\beta(-k,k')$ is zero for $k^2=1$. Thus we find
$T=\Phi+{\cal O}(\widetilde{\lambda}^3)$ which is actually
consistent with the energy matching\footnote{We note that in BCFT
it is trivial to obtain the rolling tachyon solution with
$\sinh(t)$ from the one with $\cosh(t)$ by the transformation $x^0
\rightarrow x^0 + i\pi, \widetilde{\lambda} \rightarrow i
\widetilde{\lambda}$ in the superstring case \cite{originalroll}.
Thus if we assume the analyticity and the consistency for the
$\cosh(t)$ solution, it is obvious that the field redefinition is
also consistent for the $\sinh(t)$ solution. }. To consider higher
order in $\widetilde{\lambda}$ is an interesting question. We can
instead consider a constant tachyon. 
Note that the field $\tilde{T}$ in which the tachyon potential is
the standard BSFT potential $(1+\tilde{T})e^{-\tilde{T}}$
is mapped to $T$ by $T=1-e^{-\tilde{T}}$ \cite{Coletti}.
For $\tilde{T}=c$, where $c$
is a constant, 
the field redefinition (\ref{fieldredef}) 
gives $\Phi=c-\frac{1}{3} c^2 +{\cal O}(c^3)$. The
BSFT action evaluated for this $T$ is 
$(1+\tilde{T})e^{-\tilde{T}} = 1-\f{1}{2} c^2+\f{1}{3}c^3 +{\cal O}(c^4)$, 
which is indeed equals to
$1-\f{1}{2}\Phi^2$. Thus we find the cubic term in the usual BSFT
tachyon potential is eliminated by the field redefinition
explicitly.

Finally, we will briefly comment on
a subtlety related to the rolling tachyon solution
in the BSFT.
In the BSFT action (\ref{bsftac})
we drop total divergence terms, for example
a term linear in $T$,
although the total divergence terms in the Lagrangian
can contribute to the partition function $Z$ for the
rolling tachyon solution.
(This is because
the partial integration can not be done
due to the divergence of the integration of time.)
Since the pressure
and the $Z$ are same for the D$p$-brane,
the linear term indeed appeared in the pressure
of the rolling tachyon \cite{originalroll}.
On the other hand, the energy and pressure
are local quantities, for which the
total divergence terms can not be contributed.\footnote{
Here we defined the energy and the pressure
through the coupling to the metric
as in \cite{LNT}.
In \cite{LNT} the zero mode integral is
factorized and covariantized by multiplying
the volume factor $\sqrt{g}$.
More properly, we should multiply $1/g_s$ instead of $\sqrt{g}$.
Since the closed string vertex operators for graviton and
dilaton are mixed in the Einstein flame,
we should add the contribution from the
differentiation of the dilaton
other than the graviton in order to obtain
the energy-momentum tensor.
Then the result is same by multiplying the $\sqrt{g}$ term
except an overall constant.}
Thus it seems something wrong with the BSFT action (\ref{bsftac}).
However,
we can assume the term linear in T which contains, for example,
the world volume scalar curvature for computing the
energy-momentum tensor.
In the flat space, such term does not contribute to the BSFT
action and the energy
but changes the pressure of the
homogeneous rolling tachyon solution.
(In the Noether current method, if it would be available,
this change of the
energy-momentum tensor may correspond
to the usual ambiguity of the
definition of the Noether current.)
Therefore it is reasonable
to drop the total divergence term in (\ref{bsftac}).

\section{Scaling Limit (Closed String Decoupling Limit)}

\subsection{Large $N$ Scaling limit}

As is well-known, the $c=1$ matrix model originally was come up
with as a method of discretization of the world-sheet
\cite{twodmat}. In that context, we start with a matrix model with
a cubic $\Phi^3$ or quartic interaction $\Phi^4$ and then take the
so called double scaling limit. This limit infinitely magnifies
the region near the top of the potential and thus the model
becomes equivalent to the quadratic matrix model, i.e.
(\ref{cmat}).

One may wonder if this kind of limiting procedure may be relevant
to the present interpretation from the viewpoint of open-closed
duality using unstable D$0$-branes. The open string tachyon
potentials in string field theories typically include such a cubic
$T^3$ or quartic term $T^4$. Thus one may speculate that the
string field theory action originally includes such an interaction
term and it will be simplified after a limiting process such as
the decoupling limit.

However, this naive scenario is not actually relevant in our
open-closed duality as long as we consider unstable D$0$-branes in
the 2d string theory defined by the conventional Liouville theory.
This is because the BSFT action already itself is equivalent to
the quadratic action via a field redefinition as we have shown
before. Thus we do not have to take any scaling limit to make the
action simple. In this interpretation, we inevitably encounter the
constraint\footnote{ It seems that the ``phase space'' in the
original variable is infinite dimensional due to the terms
including higher derivatives and the constraint which corresponds
to (\ref{cons}) is ambiguous. However, it is two dimensional
according to our claim which implies that the BSFT action depends
on only two independent (non-local) functions of the $\frac{d^n
T}{d t^n}$. Thus the constraint in the ``phase space'' is
unambiguous. } (\ref{cons}) when we try to relate the tachyon $T$
to the matrix $\Phi$ by the field redefinition because the energy
of the classical solutions are bounded from below. This classical
constraint can be interpreted as the infinitely many fermions
filling the energy level below $E=-\mu$. Of course, this coincides
with the matrix model description of the ZZ-branes proposed in
\cite{KMS}.

Although this argument for the existence of the infinitely many
fermions, which is responsible for the closed string excitations,
is obviously very natural, it is highly nontrivial to show it from
the BSFT point of view. One way to see this is to consider a
scaling limit of the tachyon action such that we can forget about
the energy bound. We will see also that the closed string is
decoupled from the D-brane action in this limit.

First, consider putting $N$ ZZ-branes in the 2d string background
specified by the coupling constant $g_s=1/\mu$. The energy level
of the ground state of the fermion action is a function of $N$ and
$g_s$ denoted by $\tilde{\mu}(N,g_s)$. Then we take the scaling
limit $N \rightarrow \infty$ and $g_s=1/\mu \rightarrow 0$ with
$\tilde{\mu}(N,g_s)$ fixed finite. In the limit, we can forget
about the energy bound because of $\mu \rightarrow \infty$ and the
BSFT action is simply given by the double scaled matrix model
action with the Fermi level $-\tilde{\mu}$. This also implies the
decoupling of the closed string since the the closed string
excitations are the fluctuations of the fermions with the energy
below $-\mu \rightarrow -\infty$, which cannot interact with the
finite energy fermions.

Therefore, we can obtain the matrix model action from the BSFT
action of ZZ-branes by taking this rather trivial scaling limit.
Using the known relation between the matrix model and 2d string,
it is this scaling limit that we should take in order to describe
the 2d string non-perturbatively by the theory on the D-branes.

It may be also useful to compare our interpretation with the
 AdS/CFT correspondence \cite{ads}, which is the most
well-studied example of open-closed duality. It sounds reasonable
to relate the scaling limit explained just before with the
near-horizon limit in the AdS/CFT. However, we should note that in
our example of 2d string theory, taking the limit only changes the
background trivially. We can more properly say that it is an
analogue of changing radius of the AdS background
by adding D$3$-branes
as can be
imagined from the matrix model with a harmonic potential
\cite{matym} for half BPS states in $N=4$ super Yang-Mills theory.

In this way we have observed that the standard double scaling
limit in $c=1$ matrix model is not relevant in our analysis of
open-closed duality. Nevertheless, there is an open possibility
that there exists an unknown (possibly highly curved) background
in 2d string theory or critical string theory which is exactly
described by the matrix model before taking the double scaling
limit. If this is true, the conventional 2d string theory can be
thought of as a `near horizon limit' of this new string theory in
the true sense.

\subsection{Back-reaction of ZZ-brane}

To check previous holographic interpretation of 2d string theory
in terms of ZZ-branes, let us compute\footnote{One of the
authors TT is very grateful to Andrew Strominger for very useful
discussions on which some of the results in this subsection are
based.} the back-reaction\footnote{See also \cite{MV,Xi} for
related computations.} when we put a ZZ-brane just on top of the
Fermi level. If our interpretation is correct it should shift the
value of the Fermi level by $\delta\mu =\f{1}{\rho(\mu)}\sim
\f{1}{\log\mu}$, where $\rho(\mu)$ is the density of fermionic
state. Here we use the same notation and convention of (bosonic)
Liouville theory as those in section (3.1.2).

In the quantum Liouville theory, the wave function $\psi(\phi)$
of zero-mode is given by
\be
\psi(\phi)=e^{iP\phi}+S(P)e^{-iP\phi},
\ee
where the phase
\be
S(P)=-e^{2i\delta(P)}=-\mu^{-iP/2}\f{\Gamma(iP)}{\Gamma(-iP)},
\ee
represents the reflection due to the Liouville
potential\footnote{Here $\mu$
denotes the renormalized cosmological
constant $\mu=\pi\gamma(b^2)\mu_0$.}.
When we consider the corresponding operator,
we have to multiply the
Liouville dressing $e^{\f{Q}{2}\phi}$.

The bulk one point function on the disk for a ZZ-brane can be
computed as \cite{ZZ} (setting $b=1$) \be \Psi(P)=\la
e^{i(P+\f{Q}{2})\phi}\lb =\f{2}{\s{\pi}}i \sinh(\pi
P)\mu^{-iP/2}\f{\Gamma(iP)}{\Gamma(-iP)} =\f{2}{\s{\pi}}i\sinh(\pi
P)e^{i\delta(P)}. \ee In other worlds, the boundary state of ZZ
brane can be written as follows \ba |B\lb_{ZZ}
&=&\int_{-\infty}^{\infty} dP~ \Psi(-P)|P\lb \\
&=&\int_{-\infty}^{\infty}
dP~ \Psi(-P)(e^{iP\hat{\phi}}+S(P)e^{-iP\hat{\phi}})|0\lb.
\ea

The fermion put just on top of the Fermi level is equivalent to
the unstable D$0$-brane decaying with the maximal amount of the
rolling tachyon \cite{originalroll}, so called $\lambda=1/2$
brane. This brane corresponds to an array of D-instantons at
$t=2\pi i(n+\f12) \ (n\in {\bf Z})$ along the imaginary time
\cite{originalroll,rollhalf}.
The closed string field $|C\lb$ (only massless
`tachyon' exists in our two dimensional string) is induced by the
presence of these D-branes following $|C\lb
=\f{1}{L_{0}+\bar{L}_{0}}|B\lb$ at the linear level.

We find the massless scalar field $\vp(t,\phi)$ sourced by the
$\lambda=1/2$ brane \be \vp(t,\phi)=\int dEdP\f{i\sinh(\pi
P)e^{-i\delta(P)}} {E^2-P^2-i\ep}\cdot e^{iEt} \cdot
(\sum_{n=-\infty}^{n=\infty}e^{iE(n+\f12)a})
\cdot(e^{iP\phi}+S(P)e^{-iP\phi}), \ee where we employed the
Feynman propagator. The constant $a$ should finally be equal to
$i$. To perform a sensible analytical continuation, we can first
rotate as $E=iP_{0}, t=-ix_{0}$. Next we do the integration and
finally re-rotate $x_{0}=it$ again. Then we get the expression \be
\vp(t,\phi) =i\int_{-\infty}^{\infty}\f{\cos(Pt)}{P}
(e^{-i\delta(P)}e^{ip\phi}-e^{i\delta(P)}e^{-ip\phi}). \ee Using
the formula of Bessel function (for $y>0$) \be
J_{0}(y)=\f{1}{2\pi}
\int_{-\infty}^{\infty}ds\f{\Gamma(-is)}{\Gamma(1+is)}(\f{y}{2})^{2is},
\ee we get \be
\vp(t,\phi)=-4\pi\left(J_{0}(2\mu^{\f14}e^{\f{\phi+t}{2}})+
J_{0}(2\mu^{\f14}e^{\f{\phi-t}{2}})\right). \ee

For example, in the weak coupling region $\phi\to -\infty$,
the expression is approximated by
\be
\vp(t,\phi)\sim -4\pi\left(2
-\s{\mu}e^{\phi+t}-\s{\mu}e^{\phi-t}\right).
\ee

Thus at an early time we find an incoming wave (toward the
strongly coupled region), while at late time the outgoing wave is
dominant. This is consistent with the fact that the brane creation
and annihilation take place in the strongly coupled region
($\phi\to \infty$), where the ZZ-brane is located. The constant
part shifts the cosmological constant $\mu$ by $\delta\mu\sim
\f{1}{\log\mu}$, as can be understood from its back-reaction to
the Liouville term $\mu \phi e^{2\phi}$ at the typical region
$\phi\sim -\f{1}{2}\log \mu$. This is exactly what we would like
to show.

\section{Implications for the BSFT in Higher Dimension}

It will be worth considering how our results in 2d string theory
are helpful to the understanding of the BSFT in higher dimension.
In general, the BSFT action includes infinitely many higher
derivative terms. This leads to many complications as those in the
initial value problem, the definition of its phase space, the
quantization of the theory and so on. In the 2d string case,
however, we have shown that the action can be greatly simplified
via a field redefinition into the quadratic one. Then we can
quantize everything very easily in the latter formulation. Notice
that even though this model is so simple and free, it includes
non-trivial dynamics like closed sting S-matrices.
Thus
the correct path-integral measure is given by
the standard measure for the matrix model variable
if our claim is correct and the BSFT is actually a string field
theory.
This implies that
the path-integral measure in the BSFT variable
should be complicated because of the Jacobian of the
field redefinition.
We might be able to find it from the contribution
from some fields other than tachyon.

A similar simplification may occur in the higher dimensional BSFT.
This will be very useful for further analysis of BSFT in future.
 For example, this may shed some light on
the problem of quantizing BSFT, which is very difficult at this
point (e.g. see \cite{bsftloop}). Although it is not an easy task
to pursue this direction, we can at least notice the following
interesting property of the BSFT action in the ordinary $26$
dimensional bosonic string or $10$ dimensional type II or type0
string theory. If we set all fields in the BSFT other than the
tachyon field $T$ to zero, then the unstable D$0$-brane action is
exactly the same as the one we studied in this paper. Thus this
truncated action is equivalent to the quadratic action.
Classically it is meaningful to consider the subset of the
configuration space such that the massless and massive fields
vanish and solve the equations of motion. Quantum mechanically, it
is not allowed to fix some fields to some values by hand in
general. Therefore it is very important to include the effects of,
in particular, the massless fields to the tachyon action. Though
this is very difficult problem as stated in the introduction, we
hope we will find some simple way to incorporate such effects into
the matrix model action in near future.

Another interesting lesson is
for the problem of the restriction of the momentum \cite{HaTe}.
The function $A(k)$ is divergent at $k^2=2,3,\ddd$
and becomes negative, for example $2<k^2<5/2$.
Thus the $T(k)$ is not a good coordinate for $2<k^2$.
However, in the matrix model side,
there is no momentum restriction.
This suggests that
we can extend the BSFT variable $T(k)$
to  $2<k^2$ by proper field redefinition
around $k^2=2$
although $T(k)$ is not valid for $2<k^2$
even in the higher dimensional BSFT.

\section{Conclusion}

In this paper we have argued that the BSFT action for unstable
D$0$-branes in 2d string theory is classically equivalent to the
double scaled $c=1$ matrix model action (i.e. quadratic action)
via a field redefinition. This gives a simple derivation of the
$c=1$ matrix model action from the open string field theory of
unstable D$0$-branes. We have given strong evidences for this claim.
First, we showed that the S-matrices of open string tachyons on
the D$0$-branes are trivial. Also we have explicitly constructed a
smooth field redefinition up to the cubic term in the BSFT action.
To understand this equivalence from the viewpoint of open-closed
duality, we also pointed out a scaling limit, which can be
regarded as a closed string decoupling limit. Finally we discussed
how our results in 2d string can be generalized to the BSFT in
higher dimension.

An interesting point in the equivalence is that
the field redefinition is smooth, but
non-local in time.
This can be seen from the fact
that the field redefinition (\ref{fieldredef})
contains non-polynomial functions of
the time derivatives.
Although we do not have
any physical interpretation of the non-locality,
it may be related to some interesting phenomena
and worth investigating.

There are several interesting future directions. In this paper we
only discussed BSFT among open string field theories. It would be
intriguing to study what is the precise relation between $c=1$
matrix model and the cubic string field theory\footnote{Refer to
\cite{GRF} for the construction of Kontsevich matrix model from
the cubic string field theory for FZZT-branes. A related argument 
for the cubic string field theory in
$c=1$ string can be found in \cite{GMM}. See also
\cite{Vafa} for the construction of similar matrix models from
D-branes in topological string.} \cite{Wittencubic}. It is also
curious to see if the similar arguments in our paper can be
applied to other matrix models such as those for the minimal
string.


\vskip6mm
\noindent
{\bf Acknowledgements}

\vskip2mm We would like to thank J. McGreevy and S.~Sugimoto for
useful discussions. ST is also grateful to J.~de Boer,
K.~Hashimoto, G.~Moore, A.~Parnachev, A.~Sinkovics, A.~Tseytlin
and J.~Yee for valuable discussions.
 TT thanks T. Asakawa, D. Gaiotto, J. L.Karczmarek,
S. Minwalla and A. Strominger for stimulating discussions.
 The work of TT was supported in part by
DOE grant DE-FG02-91ER40654.

\noindent

\appendix
\setcounter{equation}{0}

\newpage

\newcommand{\J}[4]{{\sl #1} {\bf #2} (#3) #4}
\newcommand{\andJ}[3]{{\bf #1} (#2) #3}
\newcommand{\AP}{Ann.\ Phys.\ (N.Y.)}
\newcommand{\MPL}{Mod.\ Phys.\ Lett.}
\newcommand{\NP}{Nucl.\ Phys.}
\newcommand{\PL}{Phys.\ Lett.}
\newcommand{\PR}{ Phys.\ Rev.}
\newcommand{\PRL}{Phys.\ Rev.\ Lett.}
\newcommand{\PTP}{Prog.\ Theor.\ Phys.}
\newcommand{\hep}[1]{{\tt hep-th/{#1}}}

\end{document}